# Higher-order exceptional ring semimetal with real hinge states in phononic crystals

Yejian Hu[1†], Zhenhang Pu[1†], Xiangru Chen[1], Yuxiang Xi[1], Jiuyang Lu[1], Weiyin Deng[1*], Manzhu Ke[1*], and Zhengyou Liu[1,2*]

[1]Key Laboratory of Artificial Micro- and Nanostructures of Ministry of Education and School of Physics and Technology, Wuhan University, Wuhan 430072, China
[2]Institute for Advanced Studies, Wuhan University, Wuhan 430072, China

†These authors contributed equally to this work.
*Corresponding author.
Emails: dengwy@whu.edu.cn; mzke@whu.edu.cn; zyliu@whu.edu.cn

Non-Hermitian topological phase, with the novel concepts such as exceptional points and skin effect, has opened up a new paradigm beyond Hermitian topological physics. Exceptional ring semimetal, featured by a stable ring of exceptional points in three dimensions, exhibits first-order topological properties, including topological surface states and surface-dependent skin effect. Nevertheless, despite extensive research on Hermitian higher-order insulators and semimetals, higher-order exceptional ring semimetal is just emerging. Here, we report the first realization of a higher-order Weyl exceptional ring semimetal in a three-dimensional lossy phononic crystal. The non-Hermitian higher-order topology reflects in the topological hinge states and hinge-dependent skin effect. Counterintuitively, the topological hinge states maintain purely real energy even under a high loss level, ensuring robust hinge-state propagation. Our findings evidence the non-Hermitian higher-order bulk-boundary correspondence of exceptional ring semimetal, and may pave the way to non-Hermitian functional acoustic devices.

Topological physics within the Hermitian paradigm has drawn substantial attention [1]. Topological semimetals are featured with nontrivial bulk band crossings [2]. For instance, the renowned Weyl semimetals exhibit doubly degenerate band crossings in three-dimensional (3D) momentum space, referred to as Weyl points (WPs). The WPs behave as monopoles of Berry flux and possess topological charges characterized by Chern numbers. In consequence, Weyl semimetals host ensured Fermi arc surface states that connect the projections of oppositely charged WPs [2]. Over the past years, the growing interest in higher-order topology arose first in insulating phases [3-5] and then in 3D semimetal phases [6-8], including the Weyl semimetals. Unlike conventional or first-order topological semimetals, higher-order Weyl semimetals are identified by topological hinge states (THSs) on their 1D boundaries [9,10]. Benefiting from the macroscope scale and the flexibility in design of classical artificial structures, higher-order Weyl semimetals have been experimentally confirmed in phononic crystals (PCs) [11-13], photonic crystals [14], and circuits [15], revealing the hierarchical nature of Weyl topology.

Recently, it is discovered that the scope of topological physics can be dramatically expanded in non-Hermitian systems. Non-Hermitian topological phases exhibit a wealth of unique phenomena, including exceptional degeneracies [16-18], spectral braiding [18-20], and skin effects [21-24]. Particularly noteworthy are non-Hermitian topological semimetals, that host exceptional degenerate points [25-26] or lines [27-29] with coalesced eigenstates [30-33], attracting significant research attention in the past year [34-37]. Among them, Weyl exceptional ring (WER) semimetals are characterized by exceptional doubly degenerate rings, termed WERs, which can be evolved from the WPs under non-Hermitian perturbations [38-41]. WER inherits the topological charge of WP behaving as a line source or sink of Berry flux, and acquires a new topological charge of spectral winding number [42-46] without Hermitian counterpart. Intriguingly, WERs are discovered to support THSs as well, bringing to light the first type of non-Hermitian higher-order topological semimetal in theory [47]. Compared with extensive research on Hermitian higher-order insulators and semimetals, investigation into non-Hermitian higher-order topological semimetal is critically lacking, especially in experiments [47,48].

Here, we report the first realization of a higher-order WER semimetal in a 3D non-Hermitian PC, which is developed from a stacked breathing Kagome lattice with loss. The PC hosts WERs with dual topological charges in terms of quantized Chern number and spectral winding number, as the first-order topologies, giving rise to Fermi arc surface states and surface-dependent skin effect, respectively. More importantly, the second-order topologies are revealed by the THSs protected by the bulk polarization, and the hinge-dependent skin effect stemming from the interplay between the Fermi arc surface states and non-Hermitian skin effect. Unlike the almost unobservable trivial hinge states, the WER-induced THSs maintain counterintuitive purely real energy, which ensures robust hinge propagation for acoustic waves against introduced loss. The theoretical, simulated, and experimental results exhibit a high degree of concordance.

We start from a three-band lattice model in Fig. 1(a), which is constructed by stacking the breathing Kagome lattice along the $z$ direction via cross-linked hoppings. The tight-binding Hamiltonian in momentum space reads

$$H = \begin{pmatrix} 0 & J_1 + J_0 e^{-i(\frac{k_x}{2}+\frac{\sqrt{3}k_y}{2})} & J_1 + J_0 e^{-ik_x} \\ J_1 + J_0 e^{i(\frac{k_x}{2}+\frac{\sqrt{3}k_y}{2})} & 0 & J_1 + J_0 e^{-i(\frac{k_x}{2}-\frac{\sqrt{3}k_y}{2})} \\ J_1 + J_0 e^{ik_x} & J_1 + J_0 e^{i(\frac{k_x}{2}-\frac{\sqrt{3}k_y}{2})} & 0 \end{pmatrix}, \quad (1)$$

with $J_1 = t_1 + 2t_2 \cos k_z$, $J_0 = t_0 + i\gamma$, and $k_i$ ($i = x, y, z$) are the dimensionless wave vectors. Loss is added in the intercell hopping $t_0 + i\gamma$ and can bring the non-Hermiticity, and other hoppings are Hermitian. For $\gamma = 0$, the system is Hermitian with real energies and hosts four WPs between the first two bands, as shown in the left panel of Fig. 1(b). These WPs reside at the corners of the $k_z = \pm k_{\text{WP}}$ planes ($k_{\text{WP}} = \text{acos}[(t_0 - t_1)/(2t_2)]$) in the first Brillouin zone (BZ), carrying topological charges of Chern numbers $\pm 1$. When increasing $\gamma$, the non-Hermiticity is introduced, and each WP will evolve into a WER located at the $k_z = \pm k_{\text{WER}}$ planes ($k_{\text{WER}} = \text{acos}[(\sqrt{t_0^2 + \gamma^2} - |t_1|)/|2t_2|]$), as shown in the right panel of Fig. 1(b). The local dispersion around a WER is visible in Fig. 1(c), the simultaneous degeneracy of the real and imaginary parts of bands only occurs at the WER. The WERs are also the borderlines interpolating the real- and imaginary-part-degenerate surfaces in momentum space, see Supplementary S1 [49].

Now we characterize the topological properties of the WERs. Originating from the interplay between the WPs and non-Hermiticity, the WERs carry dual topological charges described by Chern numbers and spectral winding numbers. The Chern number $C$ is defined as the integration of the Berry curvature over a closed surface $S$ (yellow sphere in the top panel of Fig. 1(d))

$$C = \frac{1}{2\pi i} \oint_S \nabla_k \times \langle \tilde{u}(\mathbf{k})|\nabla_k|u(\mathbf{k})\rangle \cdot d\mathbf{S}, \tag{2}$$

where $|u(\mathbf{k})\rangle$ ($|\tilde{u}(\mathbf{k})\rangle$) is the left (right) eigenvector for the first band. Naturally, the WERs inherit the nonzero $C = \pm 1$ from the WPs, as marked by the purple and cyan colors in Fig. 1(b). Besides, the winding number $\nu$ labels the spectral topology for the first two bands along a closed path $\mathcal{L}$ encircling the WER (black circle in top panel of Fig. 1(d)), defined as

$$\nu = \frac{1}{2\pi} \sum_{i,j=1}^{2} \oint_{\mathcal{L}} d\mathbf{k} \cdot \nabla_k \arg[E_i(\mathbf{k}) - E_j(\mathbf{k})], \tag{3}$$

where $E_{i(j)}$ represents the energy of the $i$- ($j$-) th band. For example, $\nu = -1$ indicates that the first two bands wind clockwise in the complex plane and form a point gap, as exemplified in the bottom panel of Fig. 1(d). Numerical calculations show that the winding numbers of the WERs at the $k_z = \pm k_{\text{WER}}$ planes correspond to $\nu = \mp 1$. Owing to the dual topological charges, the Fermi arc surface states and surface-dependent skin effect are ensured, see details in Supplementary S2 and S3 [49].

To demonstrate the second-order topology, we regard our 3D system as a set of $k_z$-dependent 2D subsystems and investigate the second-order topological index, that is, the bulk polarization, defined based on 2D BZ

$$p_{x,y}(k_z) = \frac{1}{S} \iint_{\text{2D BZ}} -i \langle \tilde{u}(\mathbf{k})|\partial_{k_{x,y}}|u(\mathbf{k})\rangle d^2k, \tag{4}$$

where $S$ is the area of the 2D BZ. Figure 1(e) shows the evolution of the bulk polarization for the first band as a function of $k_z$, a nontrivial $(p_x, p_y) = (1/2, 1/2\sqrt{3})$ emerges for $|k_z| > k_{\text{WER}}$ within the hinge BZ and indicates the presence of THSs in this region. To verify the THSs, we demonstrate the hinge dispersion in Fig. 1(f), with its imaginary part given by the colormap. As expected, a branch of zero-energy THS dispersion emerges, which are localized at left lower hinge of the geometry in Fig. 1(a) owing to the bulk-hinge correspondence [50].

As a second-order skin effect, the hinge-dependent skin effect is further discussed in our model. Note to lose the generality, we focus on the XZ surfaces and calculate the energy spectrum of a ribbon structure with periodic boundary conditions in the $x$ and $z$ directions. As shown in Fig. 1(g), the surface states (green dots) along a route $k_z = k_x - \pi/3$ in the surface BZ form two point gaps, with the left one corresponding to Fermi arc surface states ensured by the WERs. That means the XZ surface states can accumulate along the $k_z = k_x$ direction under fully open boundary conditions and localize at $[\bar{2}112]$-directional hinges as skin modes, as shown in Fig. 1(h). The colormap represents the distribution of the XZ surface states defined as $W_s(j) = \frac{1}{N_s}\sum_{n_s}|\psi_{n_s}(j)|^2$, where $\psi_{n_s}(j)$ is the $n_s$-th right eigenstate at site $j$ and $N_s$ denotes the total number of surface states. However, due to the mirror symmetries along the $x$ and $z$ directions, such skin modes are selectively forbidden on horizontal and vertical hinges. In other words, the second-order skin effect here is hinge-dependent, as depicted by the arrows in Fig. 1(h), which is significantly different from 2D scenarios. See details in Supplementary S3 [49].

Then we implement the aforementioned model in a 3D non-Hermitian PC. Figure 2(a) gives the photographs of the 3D-printed PC sample, which is a rhombic prism containing $13 \times 13 \times 13$ unit cells with the in-plane and out-of-plane lattice constants $a = 40$ mm and $h = 45.87$ mm. In each unit cell, as illustrated in Fig. 2(b), there are three hexagonal prism cavities of height $h_0 = 20$ mm and width $d_0 = 10$ mm, which are filled with air and correspond to the sites in the lattice model. The connecting tubes between the cavities mimic the hoppings. Specifically, the staggered intralayer hoppings are introduced by cuboid tubes with $d_1 = 4.2$ mm and $d_2 = 3.5$ mm, while the interlayer hoppings are enabled by cross-linked tubes with $d_3 = 5/\sqrt{3}$ mm and $d_4 = 5$ mm. To add loss in our PC, sponge-filled rectangular holes (blue areas in Fig. 2(b), $d_5 = 4.4$ mm and $h_1 = 3.8$ mm) are delicately designed on the intercell tube walls, see Supplementary S4 for details [49]. In simulations, impendence boundaries of 1500 Pa·s/m are applied to these blue areas [51], and hard boundaries to the gray regions considering the huge acoustic impedance mismatch between the 3D-printed plastic material and air. Enlarged configuration of the sponges (black) is also provided in the inset of Fig. 2(a). It is worth noting that, to distinguish and detect the bulk, surface, and hinge dispersions in experiments, we construct the PC sample in the geometry in Fig. 1(a) where skin modes are forbidden. Simulated results of surface-dependent and hinge-dependent skin effects for acoustic waves, as the first-order and second-order skin effects, are provided in Supplementary S6 [49].

We demonstrate the acoustic WER by measuring the bulk band structure of the PC. Figure 2(c) illustrates the experimental measurement. A broadband point source (green star) is placed at the center of the PC to excite the bulk states, and the acoustic signals are recorded by a microphone inserted into the cavities. Two slices of the measured acoustic field are exemplified by the colormaps in Fig. 2(c), one can see that the wave propagation in the $x$-$y$ plane is weaker than that along the $z$ direction, which is due to the loss added in the intralayer hoppings. By Fourier transforming the 3D acoustic field, we obtain the band structure of the PC. Focus on the $k_z = -k_{\text{WER}}$ plane, we display the bulk dispersion in Fig. 2(d), where two WERs are clearly observed at the frequency $f = 7.74$ kHz. The gray lines and colormap represent the simulated and measured results, respectively. The corresponding high symmetry route is depicted by the orange line in Fig. 2(e), which passes through the WERs at four exceptional points round $K_1$ and $K_1'$. These crossing points are highlighted by purple and cyan spheres in Fig. 2(d). For the region inside the WERs, the first two bands are relatively flat but not degenerate, consistent with the theoretical prediction. In Fig. 2(f), we further display the isofrequency contour at the WER frequency, where the triangle-like shape of the WERs and the relatively flat bands are commendably verified.

As a significant signal of WER semimetals, the topologically ensured Fermi arc surface states are validated in our PC. Figure 3(a) gives the experimental measurements. The source (green star) is centered at the front XZ surface, and the Fermi arc surface states are experimentally accessible from the Fourier transformation of the surface acoustic field (colormap). The corresponding surface BZ is shown in Fig. 3(b), where the projected WERs are restricted to the purple and cyan lines. Owing to the first-order topological nature of the WERs, two open arcs of topological surface states emerge in the isofrequency contour in Fig. 3(c), connecting the projected WERs with opposite Chern numbers. The dashed lines label specific $k_z = -2/3, -1/2, -1/3\ \pi/h$, with the corresponding surface dispersions presented in Fig. 3(d). The gray and green lines denote simulated bulk and surface states, respectively. Note that, due to the zero $k_z$-dependent Chern number, the surface states do not touch the bulk states for some $k_z$, e.g., in the left panel of Fig. 3(d), but are gapless as a whole.

To validate the second-order topology in our PC, experimental measurement for the THSs is performed in Fig. 4(a). One can see from the colormap that, with respect to a source (green star) centered at the left hinge, the acoustic wave is localized at the hinge and propagates along the $z$ direction. Obtained by Fourier transforming the 1D acoustic field at the hinge, the hinge dispersion of the PC is given by the colormap in Fig. 4(b). Because of the nontrivial bulk polarization, a branch of THS dispersion emerges at around 8.34 kHz in the region of $|k_z| > k_{\mathrm{WER}}$ as expected, identifying our PC as a second-order WER semimetal. The dots display the simulated result for comparison, agreeing with the experimental data well.

Besides the THSs, there are two branches of trivial hinge states at around 7.24 and 8.90 kHz, which are localized at the right hinge of the PC. To be more explicit, we display the simulated hinge dispersion in Fig. 4(c) with the imaginary parts labeled by colormap. One can see that the trivial hinge states have larger imaginary parts of frequencies, implying that the hinge wave propagation is influenced by the system loss. On the contrary, the THSs have negligible imaginary parts of frequencies, revealing their robustness against non-Hermitian perturbations. This fact can be confirmed by the response spectra for acoustic wave propagating along different hinges of our sample, as shown in Fig. 4(d), measured by exciting and probing two adjacent cavities on the corresponding hinges. The response curve for the left hinge (blue line) reaches a strong peak in the blue region, representing the THSs, while the curve for the right hinge (red line) has a much weaker intensity. See experimental data for trivial hinge states in

Supplementary S7 [49].

In summary, we have realized a second-order WER semimetal in a 3D PC, where the non-Hermiticity is introduced by designed loss. The acoustic WERs carry dual topological charges that result in Fermi arc surface states and surface-dependent skin effect, and the second-order topologies are reflected by robust 1D THSs and hinge-dependent skin effect. Notably, given the unavoidable intrinsic loss in natural and engineered systems, the WER-induced real hinge state holds significant potential for applications in topological waveguides and sensors. Our work broadens the insight to the interplay between topological semimetal and non-Hermiticity, and may pave the way for the development of non-Hermitian acoustic devices.

**Acknowledgements**

This work is supported by the National Key R&D Program of China (Nos. 2022YFA1404500, 2022YFA1404900), National Natural Science Foundation of China (Nos. 12074128, 12222405, 12374409, 12574024, 12504250), Hubei Provincial Natural Science Foundation of China (No. 2025AFB051), China Postdoctoral Science Foundation (No. 2024M752453), and Postdoctoral Fellowship Program of CPSF (No. GZB20240575).

**References**

[1] C.-K. Chiu, J. C. Y. Teo, A. P. Schnyder, and S. Ryu, Classification of topological quantum matter with symmetries, Rev. Mod. Phys. **88**, 035005 (2016).

[2] N. P. Armitage, E. J. Mele, and A. Vishwanath, Weyl and Dirac semimetals in three-dimensional Solids, Rev. Mod. Phys. **90**, 015001 (2018).

[3] M. Ezawa, Higher-order topological insulators and semimetals on the Breathing kagome and pyrochlore lattices, Phys. Rev. Lett. **120**, 026801 (2018).

[4] W. A. Benalcazar, B. A. Bernevig, and T. L. Hughes, Quantized electric multipole insulators, Science **357**, 61 (2017).

[5] H. Zhong et al., Observation of nonlinear fractal higher order topological insulator, Light Sci. Appl. **13**, 264 (2024).

[6] B. J. Wieder, Z. Wang, J. Cano, X. Dai, L. M. Schoop, B. Bradlyn, and B. A. Bernevig, Strong and fragile topological Dirac semimetals with higher-order Fermi arcs, Nat. Commun. **11**, 627 (2020).

[7] Z. Wang, D. Liu, H. T. Teo, Q. Wang, H. Xue, and B. Zhang, Higher-order Dirac

semimetal in a photonic crystal, Phys. Rev. B **105**, L060101 (2022).

[8] Q. Ma, Z. Pu, L. Ye, J. Lu, X. Huang, M. Ke, H. He, W. Deng, and Z. Liu, Observation of higher-order nodal-line semimetal in phononic crystals, Phys. Rev. Lett. **132**, 066601 (2024).

[9] H.-X. Wang, Z.-K. Lin, B. Jiang, G.-Y. Guo, and J.-H. Jiang, Higher-order Weyl semimetals, Phys. Rev. Lett. **125**, 146401 (2020).

[10] S. A. A. Ghorashi, T. Li, and T. L. Hughes, Higher-order Weyl semimetals, Phys. Rev. Lett. **125**, 266804 (2020).

[11] Q. Wei, X. Zhang, W. Deng, J. Lu, X. Huang, M. Yan, G. Chen, Z. Liu, and S. Jia, Higher-order topological semimetal in acoustic crystals, Nat. Mater. **20**, 812 (2021).

[12] L. Luo, H.-X. Wang, Z.-K. Lin, B. Jiang, Y. Wu, F. Li, and J.-H. Jiang, Observation of a phononic higher-order Weyl semimetal, Nat. Mater. **20**, 794 (2021).

[13] Z. Pu, H. He, L. Luo, Q. Ma, L. Ye, M. Ke, and Z. Liu, Acoustic higher-order Weyl semimetal with bound hinge states in the continuum, Phys. Rev. Lett. **130**, 116103 (2023).

[14] Y. Pan et al., Real higher-order Weyl photonic crystal, Nat. Commun. **14**, 6636 (2023).

[15] L. Song, H. Yang, Y. Cao, and P. Yan, Square-root higher-order Weyl semimetals, Nat. Commun. **13**, 5601 (2022).

[16] E. J. Bergholtz, J. C. Budich, and F. K. Kunst, Exceptional topology of non-Hermitian systems, Rev. Mod. Phys. **93**, 015005 (2021).

[17] Y. Wu, Y. Wang, X. Ye, W. Liu, Z. Niu, C. K. Duan, Y. Wang, X. Rong, and J. Du, Third-order exceptional line in a nitrogen-vacancy spin system, Nat. Nanotechnol. **19**, 160 (2024).

[18] H. Hu and E. Zhao, Knots and non-Hermitian Bloch bands, Phys. Rev. Lett. **126**, 010401 (2021).

[19] K. Wang, A. Dutt, C. C. Wojcik, and S. Fan, Topological complex-energy braiding of non-Hermitian bands, Nature **598**, 59 (2021).

[20] Z. Rao, C. Meng, Y. Han, L. Zhu, K. Ding, and Z. An, Braiding reflectionless states in non-Hermitian magnonics, Nat. Phys. **20**, 1904 (2024).

[21] L. Zhang et al., Acoustic non-Hermitian skin effect from twisted winding topology, Nat. Commun. **12**, 6297 (2021).

[22] Z. Pang, B. T. T. Wong, J. Hu, and Y. Yang, Synthetic non-Abelian gauge fields for non-Hermitian systems, Phys. Rev. Lett. **132**, 043804 (2024).

[23] Z. Li, L.-W. Wang, X. Wang, Z.-K. Lin, G. Ma, and J.-H. Jiang, Observation of dynamic non-Hermitian skin effects, Nat. Commun. **15**, 6544 (2024).

[24] K. Zhang, Z. Yang, and C. Fang, Universal non-Hermitian skin effect in two and higher dimensions, Nat. Commun. **13**, 2496 (2022).

[25] Y. Ashida, Z. Gong, and M. Ueda, Non-Hermitian physics, Adv. Phys. **69**, 249 (2020).

[26] W. D. Heiss, The physics of exceptional points, J. Phys. A Math. Theor. **45**, 444016 (2012).

[27] W. Tang, X. Jiang, K. Ding, Y.-X. Xiao, Z.-Q. Zhang, C. T. Chan, and G. Ma, Exceptional nexus with a hybrid topological invariant, Science **370**, 1077 (2020).

[28] W. Tang, K. Ding, and G. Ma, Realization and topological properties of third-order exceptional lines embedded in exceptional surfaces, Nat. Commun. **14**, 6660 (2023).

[29] C. Wang, N. Li, J. Xie, C. Ding, Z. Ji, L. Xiao, S. Jia, B. Yan, Y. Hu, and Y. Zhao, Exceptional nexus in Bose-Einstein condensates with collective dissipation, Phys. Rev. Lett. **132**, 253401 (2024).

[30] Q. Zhou, J. Wu, Z. Pu, J. Lu, X. Huang, W. Deng, M. Ke, and Z. Liu, Observation of geometry-dependent skin effect in non-Hermitian phononic crystals with exceptional points, Nat. Commun. **14**, 4569 (2023).

[31] M. Król et al., Annihilation of exceptional points from different Dirac valleys in a 2D photonic system, Nat. Commun. **13**, 5340 (2022).

[32] H. Wang, J. Ruan, and H. Zhang, Non-Hermitian nodal-line semimetals with an anomalous bulk-boundary correspondence, Phys. Rev. B **99**, 075130 (2019).

[33] Z. Yang and J. Hu, Non-Hermitian Hopf-link exceptional line semimetals, Phys. Rev. B **99**, 081102 (2019).

[34] Y. Wu, D. Zhu, Y. Wang, X. Rong, and J. Du, Experimental observation of Dirac exceptional points, Phys. Rev. Lett. **134**, 153601 (2025).

[35] Z. Lei and Y. Deng, Topological dynamics and correspondences in composite exceptional rings, Commun. Phys. **8**, 67 (2025).

[36] Z. Na, X. Zhang, and J. Feng, Multiple exceptional rings in one-dimensional perovskite photonic crystals, Phys. Rev. Lett. **135**, 083802 (2025).

[37] Y. Hu, J. Wu, P. Ye, W. Deng, J. Lu, X. Huang, Z. Wang, M. Ke, and Z. Liu, Acoustic exceptional line semimetal, Phys. Rev. Lett. **134**, 116606 (2025).

[38] A. Cerjan, M. Xiao, L. Yuan, and S. Fan, Effects of non-Hermitian perturbations on Weyl hamiltonians with arbitrary topological charges, Phys. Rev. B **97**, 075128 (2018).

[39] Y. Xu, S.-T. Wang, and L.-M. Duan, Weyl exceptional rings in a three-dimensional dissipative cold atomic gas, Phys. Rev. Lett. **118**, 045701 (2017).

[40] J. Liu, Z. Li, Z.-G. Chen, W. Tang, A. Chen, B. Liang, G. Ma, and J.-C. Cheng, Experimental realization of Weyl exceptional rings in a synthetic three-dimensional non-Hermitian phononic crystal, Phys. Rev. Lett. **129**, 084301 (2022).

[41] A. Cerjan, S. Huang, M. Wang, K. P. Chen, Y. Chong, and M. C. Rechtsman, Experimental realization of a Weyl exceptional ring, Nat. Photonics **13**, 623 (2019).

[42] K. Zhang, Z. Yang, and C. Fang, Correspondence between winding numbers and skin modes in non-Hermitian systems, Phys. Rev. Lett. **125**, 126402 (2020).

[43] Z. Gong, Y. Ashida, K. Kawabata, K. Takasan, S. Higashikawa, and M. Ueda, Topological phases of non-Hermitian systems, Phys. Rev. X **8**, 031079 (2018).

[44] N. Okuma, K. Kawabata, K. Shiozaki, and M. Sato, Topological origin of non-Hermitian skin effects, Phys. Rev. Lett. **124**, 086801 (2020).

[45] D. S. Borgnia, A. J. Kruchkov, and R.-J. Slager, Non-Hermitian boundary modes and topology, Phys. Rev. Lett. **124**, 056802 (2020).

[46] Z. Yang, A. P. Schnyder, J. Hu, and C.-K. Chiu, Fermion doubling theorems in two-dimensional non-Hermitian systems for Fermi points and exceptional points, Phys. Rev. Lett. **126**, 086401 (2021).

[47] T. Liu, J. J. He, Z. Yang, and F. Nori, Higher-order Weyl-exceptional-ring semimetals, Phys. Rev. Lett. **127**, 196801 (2021).

[48] S. A. A. Ghorashi, T. Li, M. Sato, and T. L. Hughes, Non-Hermitian Higher-Order Dirac Semimetals, Phys. Rev. B 104, L161116 (2021).

[49] See Supplemental Material for S1. Spectral degeneracy, S2. Fermi arc surface states, S3. Non-Hermitian skin effects, S4. Designed loss in PC, S5. Spectral winding number of the acoustic WERs, S6. Non-Hermitian skin effects for acoustic waves, S7. Observation of the trivial hinge states, S8. Methods, which includes Refs. [24, 51].

[50] H. Xue, Y. Yang, F. Gao, Y. Chong, and B. Zhang, Acoustic higher-order topological insulator on a Kagome lattice, Nat. Mater. **18**, 108 (2019).

[51] Y. Yang, H. Jia, Y. Bi, H. Zhao, and J. Yang, Experimental demonstration of an acoustic asymmetric diffraction grating based on passive parity-time-symmetric medium, Phys. Rev. Appl. **12**, 034040 (2019).

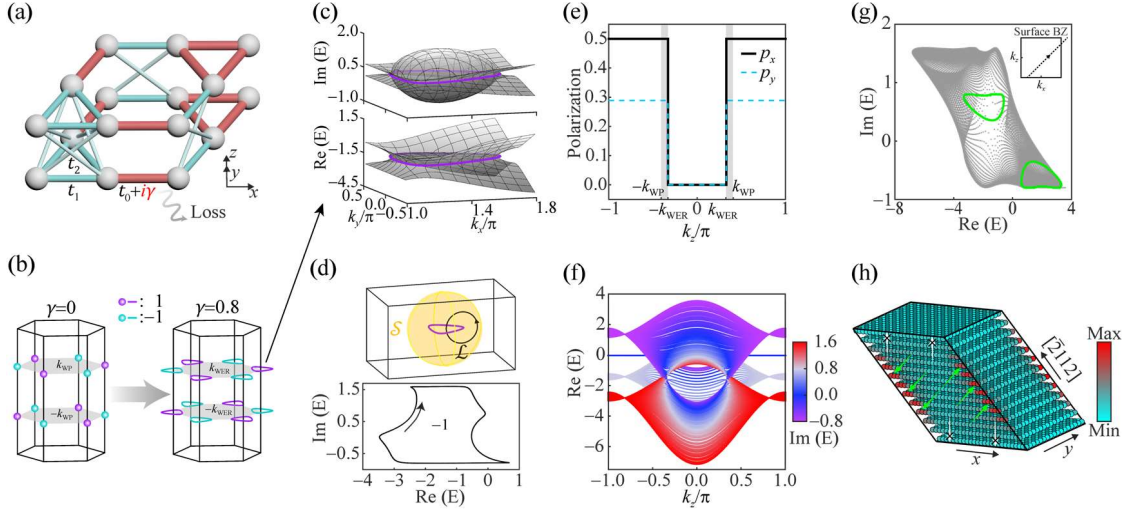

FIG. 1. Second-order WER semimetal for a 3D stacked Kagome lattice. (a) Schematic of the 3D lattice in a rhombic prism geometry, with three sublattices in a unit cell. Loss is added in the intercell hopping $t_0 + i\gamma$, while $t_1$ and $t_2$ are Hermitian. (b) Distributions of the WPs (left panel) and WERs (right panel) in the BZ. The topological charges of Chern number are indicated by the purple ($+1$) and cyan ($-1$) colors. (c) Real and imaginary parts of the local dispersion around one WER in (b), as marked by the black arrow. (d) Top panel: a WER enclosed by a surface $\mathcal{S}$ (yellow sphere) and encircled by a path $\mathcal{L}$ (black circle). Bottom panel: complex energy spectrum for the first two bands along $\mathcal{L}$. (e) Bulk polarization $(p_x, p_y)$ of the lowest band as a function of $k_z$. (f) Real part of the hinge dispersion along the $k_z$ direction, with the imaginary part given by the colormap. (g) Complex energy spectrum for the XZ surface states, along the route $k_z = k_x - \pi/3$ in the surface BZ (dashed line in the inset). The gray and green dots represent the bulk and surface states, respectively. (h) Hinge-dependent skin effect, where the XZ surface states selectively accumulate to $[\bar{2}112]$-directional hinges. The parameters are chosen as $t_0 = -1.385$, $t_1 = -1$, $t_2 = -0.6$ and $\gamma = 0.8$.

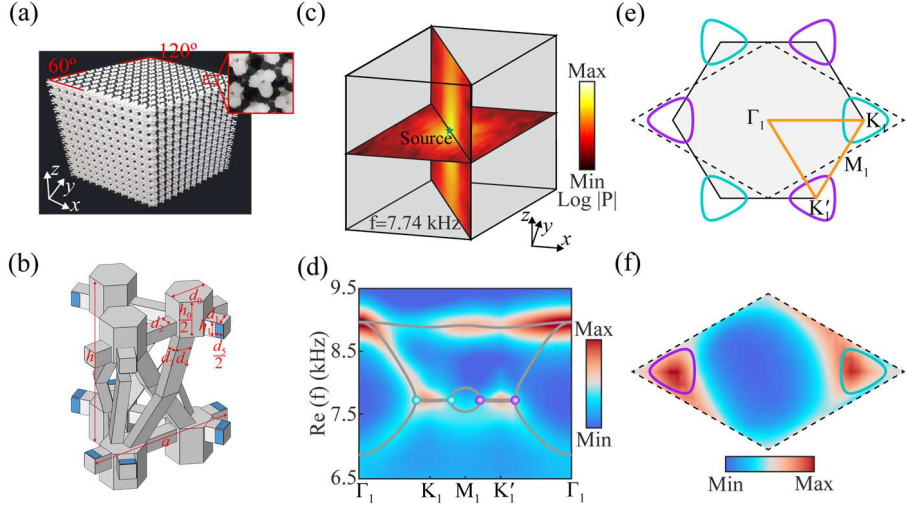

FIG. 2. Acoustic realization of the second-order WER semimetal and observation of the WERs. (a) Photographs of the non-Hermitian PC sample. The inset gives the enlarged top view. (b) Schematic of the PC unit cell. The blue areas represent designed rectangular holes, which are sealed with sponges and bring loss. (c) Experiment for the WER observation. The colormaps exemplify two slices of the measured acoustic field inside the PC at 7.74 kHz. (d) Measured (colormap) and simulated (gray lines) bulk dispersion along the orange high-symmetry lines depicted in (e). The colored spheres represent exceptional points in the WERs. (e) Simulated configuration of the acoustic WERs on the $k_z = -k_{\text{WER}}$ plane. (f) Measured isofrequency contour (colormap) at the WER frequency, corresponding to the dashed rhombus in (e).

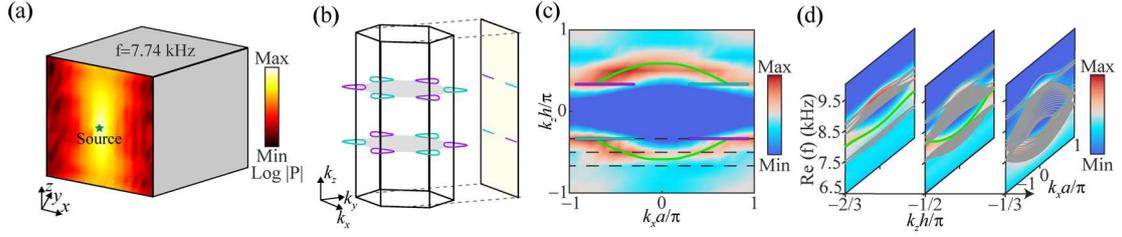

FIG. 3. Observation of the Fermi arc surface states. (a) Experiment for the surface states observation. The colormap represents the measured surface field at 7.74 kHz. (b) Projections of the WERs in the surface BZ of the XZ surface. (c) Measured (colormap) and simulated (green lines) isofrequency contour of the surface waves at the WER frequency, where the surface states connect the projections of WERs with opposite Chern numbers. (d) Measured surface dispersions (colormap) in terms of discretized $k_z$, corresponding to the black dashed lines in (c). Simulated bulk and surface states are denoted by gray and green lines, respectively.

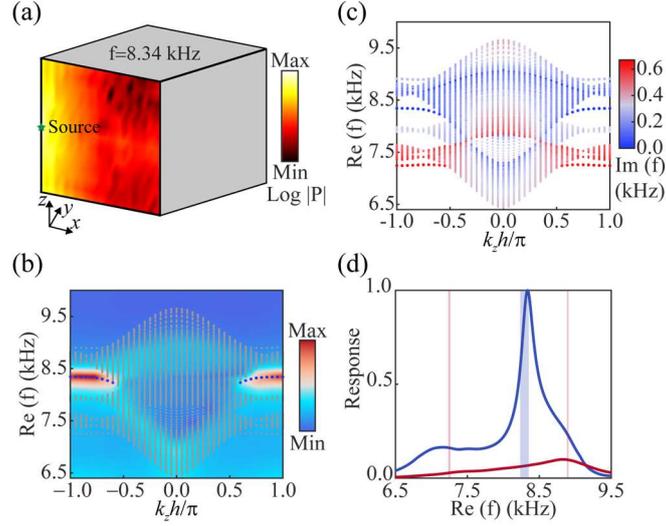

FIG. 4. Observation of the hinge states. (a) Experiment for the hinge state observation. The colormap represents the measured hinge field at 8.34 kHz. (b) Measured hinge dispersion (colormap) for the left hinge. Simulated result is denoted by gray dots and the THSs are highlighted in blue. (c) Real part of the simulated hinge dispersion of the PC, where the imaginary part is given by the colormap. The THSs around 8.34 kHz are localized at the left hinge, and the trivial hinge states around 7.24 kHz and 8.90 kHz at the right hinges. (d) Measured response spectra for the left (blue line) and right (red line) hinges of the PC. The blue and red regions label the frequency ranges of the THSs and trivial hinge states, respectively.